\documentclass[fleqn,usenatbib]{mnras}
\usepackage{newtxtext,newtxmath}
\usepackage{listings}
\usepackage{booktabs,caption}

\usepackage[T1]{fontenc}
\usepackage{longtable}
\DeclareRobustCommand{\VAN}[3]{#2}
\let\VANthebibliography\thebibliography
\def\thebibliography{\DeclareRobustCommand{\VAN}[3]{##3}\VANthebibliography}

\usepackage{graphicx}	
\usepackage{amsmath}	
\usepackage[table]{xcolor}

\newcommand{\mearth}{M$_\oplus$}

\newcommand{\kms}{\ensuremath{\rm km\,s^{-1}}}
\newcommand{\ms}{\ensuremath{\rm m\,s^{-1}}}
\newcommand{\cms}{\ensuremath{\rm cm\,s^{-1}}}

\newcommand{\teff}{T$_\mathrm{eff}$ }

\newcommand{\Feight}{F$_8$}
\newcommand{\Ic}{$\mathrm{I}_{\mathrm{c}}$}

\newcommand{\logg}{$\log g$}

\newcommand{\citeg}[1]{\citep[e.g.,][]{#1}}

\title[Stellar Granulation]{Predicting convective blueshift and radial-velocity dispersion due to granulation for FGK stars }

\author[S. Dalal et al.]{
S. Dalal,$^{1}$\thanks{E-mail: s.dalal@exeter.ac.uk}
{R. D. Haywood}$^{1}$\thanks{STFC Ernest Rutherford Fellow},
A. Mortier$^{2}$,
W.J. Chaplin$^{2}$
and N. Meunier$^{3}$
\\
$^{1}$Astrophysics Group, University of Exeter, Exeter EX4 2QL, UK\\
$^{2}$School of Physics \& Astronomy, University of Birmingham, Edgbaston, Birmingham, B15 2TT, UK\\
$^{3}$Universit\'{e} Grenoble Alpes, CNRS, IPAG, F-38000 Grenoble, France\\
}

\date{Accepted XXX. Received YYY; in original form ZZZ}

\pubyear{2015}

\begin{document}
\label{firstpage}
\pagerange{\pageref{firstpage}--\pageref{lastpage}}
\maketitle
\begin{abstract}
To detect Earth-mass planets using the Doppler method, a major obstacle is to differentiate the planetary signal from intrinsic stellar variability (e.g., pulsations, granulation, spots and plages). Convective blueshift, which results from small-scale convection at the surface of Sun-like stars, is relevant for Earth-twin detections as it exhibits Doppler noise on the order of 1 \ms. 
Here, we present a simple model for convective blueshift based on fundamental equations of stellar structure. Our model successfully matches observations of convective blueshift for FGK stars. Based on our model, we also compute the intrinsic noise floor for stellar granulation in the radial velocity observations. We find that for a given mass range, stars with higher metallicities display lower radial-velocity dispersion due to granulation, in agreement with MHD simulations. We also provide a set of formulae to predict the amplitude of radial-velocity dispersion due to granulation as a function of stellar parameters. Our work is vital in identifying the most amenable stellar targets for EPRV surveys and radial velocity follow-up programmes for TESS, CHEOPS, and the upcoming PLATO mission. 

{}

\end{abstract}

\begin{keywords}
Sun: granulation -- techniques: radial velocities -- convection -- stars: solar-type -- stars: activity 
\end{keywords}



\section{Introduction}
In the last couple of decades, more than 900 exoplanets have been discovered by the radial-velocity (RV) method\footnote{\url{https://exoplanetarchive.ipac.caltech.edu/docs/counts_detail.html}}. With the help of state-of-the-art ground-based spectrographs (e.g., see \citealt{Crass2021} and references therein), we have been pushing the detection limits towards smaller and lower mass planets (<10 \mearth). Nevertheless, we are yet to detect a planet with similar observed properties to the Earth. To achieve this goal, we must detect RV signals at a \cms\ level, which remains challenging owing to intrinsic stellar variability. This variability induces correlated RV signals with 0.1$-$100 \ms\ amplitudes that span from timescales of seconds to decades (e.g., see reviews by \citealt{Fischer2016, Dumusque2016, Cameron2018, Cegla2019, Hatzes2019, Meunier2021article} and references therein). Apart from the stellar photosphere, the RV technique is currently limited by instrumental systematics, such as wavelength calibration and drifts, and the Earth’s atmosphere, including sampling and telluric contamination. Further information can be found in Figure 1.1 of  \citealt{Crass2021}. Here we consider a radial-velocity signal which does not come from exoplanets as \textit{noise}.

Host-star activity can either mask or mimic planet-induced RV signals. There have been several controversial detections (e.g., $\alpha$ Cen B b:  \citealt{Dumusque2012, Hatzes2013, Rajpaul2016}, GJ581 d and g: \citealt{Vogt2010, Robertson2014}) where stellar activity was masquerading as planets. 

Stellar variability has timescales from seconds to years. On the timescales of the order of minutes, stellar oscillations and granulation induce RV signals with amplitudes of a few m/s. The flux imbalance caused by spots and faculae produces RV variations on rotation timescales. The suppression of convective blueshift caused by strong magnetic fields in active regions (predominantly plage) leads to RV variations on timescales of stellar rotation and magnetic cycle (See \citealt{Meunier2021article} for a review on stellar variability in RV and further references).

  \begin{figure*}
   \centering
   \includegraphics[scale=0.38]{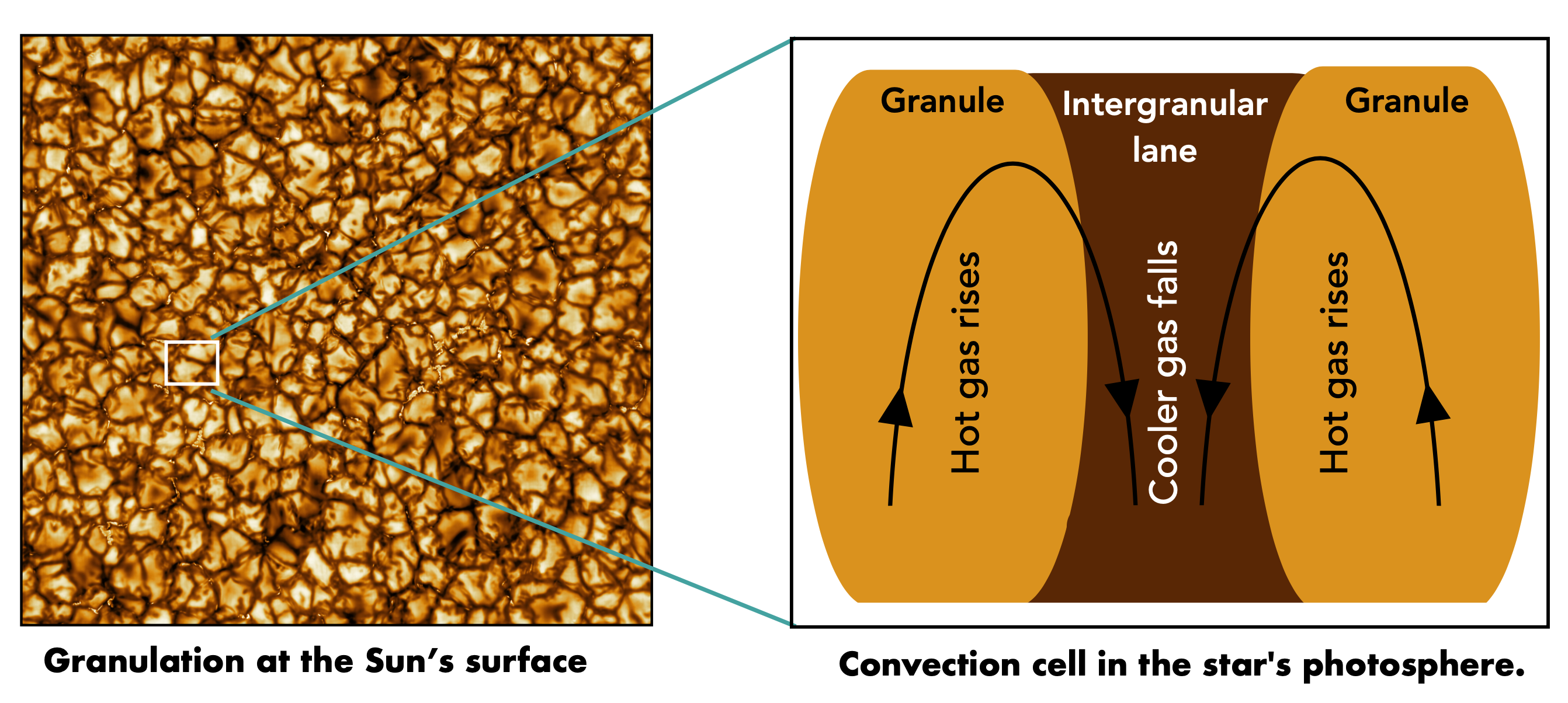}
  \vspace*{-0.20in}
   \caption{\textit{Solar granulation:}  \textbf{Left panel : } Our Sun's surface taken at 789 nm using the Daniel K. Inouye Solar Telescope (by NSO/AURA/NSF), which covers an area of $36\,500 \times 36\,500$ km (51 $\times$ 51 arcseconds). The image shows the granulation pattern, where the cell-like structures or granular cells (with dimensions of 30-1500 km) are the convection cells covering the entire surface of the Sun. Hot plasma rises in the bright cells (granules), cools down, and then falls into the dark (intergranular) lanes. \textbf{Right panel :} Illustration of the cross-section of a granular cell, spanning approximately 30-1500 km in length, where hot gas rises in the granules (orange region) and falls in the intergranular lane (dark brown region).}
              \label{Figartgranule}%
    \end{figure*}

Various methods have been developed to mitigate the effects of stellar noise. These include observational strategies that average out oscillations over time \citep{Dumusque2011,Chaplin2019}, using Gaussian Processes to model correlated stellar noise \citep{Haywood2014, Rajpaul2015, Barragan2019}, de-trending the stellar activity signal using chromospheric activity or line shape indicators\citep{Boisse2009, Robertson2014, Tuomi2014}, and building simple models of stellar surface features \citep{Lanza2010, Boisse2012}. However, stellar variability leads to the suppression of convective blueshift \citep{Meunier2010, Jeffers2014, Bauer2018, Moulla2022}. This leads to difficulty in mitigating the granulation signal with observation strategies \citep{Chaplin2019, Dumusque2011}. 

Along with mitigation, many previous studies predict the amplitude of stellar RV noise. \citet{Aigrain2012} predict the stellar RV noise of a star from a well-sampled light curve assuming a simple spot model. \citet{Saar1998} estimate the RV variation due to surface magnetic and convective activity for F, G and K stars. \citet{Bastien2013} measure the 8-hrs flicker (\Feight), i.e., the low-level photometric variations arising from granulation in the integrated light of a star occurring on timescales shorter than 8 hrs. \citet{Bastien2014} find that the \Feight\ is anti-correlated with the RV rms (root-mean-square). They conclude that the RV noise for quiet stars (i.e. photometric variability < $\sim$ 3 mmag) is driven by convective motions. More recently, \citet{Diaz2022} provide new scaling relations for stellar brightness fluctuations from K-dwarfs to giant stars (with a given temperature, surface gravity, and metallicity) for long time series of 3D stellar atmosphere models, generated with the \textsc{stagger} code \citep{Magic2013}. \citet{Zhao2022} show that photometric scaling relations from Kepler photometry could help predict variability in radial velocity due to stellar granulation and oscillation.

Being able to accurately determine the amplitude of stellar RV noise for a very large number of stars allows us to select targets with low intrinsic variability. This selection process helps to identify the most amenable targets for RV follow-up programmes for TESS, CHEOPS and the upcoming PLATO mission \citep{Rauer2014PLATO,Ricker2015TESS,Benz2021CHEOPS}. Knowing the RV contribution from stellar activity yields more reliable detections for RV planet surveys like the Terra Hunting Experiment \citep{Hall2018}.  In this paper, we predict the intrinsic noise floor for one of the stellar processes, i.e., granulation.

Sun-like stars (mid-F to late K main sequence stars) have a convective envelope in their outer layer, where the hotter bubbles of gas are buoyant and rise, whereas the cooler bubbles sink. These motions result in irregular cellular patterns at the surface, known as granulation. Figure \ref{Figartgranule} shows an observation of granulation at the Sun's surface taken by the Daniel K. Inouye Solar telescope (by NSO/AURA/NSF) and an illustration of the cross-section of a granule. The horizontal length scales for granulation in Sun range from $\sim$ 0.5 to 2 Mm (megameters) \citep{Rieutord2010}. The granulation pattern comprises granules wherein hot plasma rises up, cools off at the surface, and finally sinks back between the granules into intergranular lanes. Since the plasma in granules rises (moving towards the observer), it introduces a blueshift, contrary to the redshift introduced by the intergranular lanes. Granules typically cover a broader surface area, and they are brighter due to the hot plasma, leading to a greater contribution to the stellar flux. The imbalance between the relative contribution of hot granules and cool intergranular lanes leads to the net radial-velocity blueshift, also known as convective blueshift ($\mathcal{V}$, see review by \citealt{Dravins1982} and references within). 

Detecting granulation in stars other than our Sun is challenging, as we can not resolve the granules on their surface. However, we can detect the effect of granulation on various spectral lines. One technique that has been exploited is the third signature of granulation, i.e., the relation between the depth of spectral lines and their core position \citep{Hamilton1999}.  The physical interpretation of the differential shifts of spectral lines is as follows: as the hot plasma in granules rises, its velocity decreases with height. Since the cores of the weak lines are formed deep in the stellar photosphere, they show a larger shift in velocity compared to the strong lines. In other words, weaker spectral lines are more blueshifted than stronger lines. We can also determine the strength of convective blueshift for stars by following the third-signature scaling approach \citep{Gray2009}. Previous works employed a similar approach to obtain the differential shifts of spectral lines for the Sun \citep{Dravins1982} and other stars \citeg{Allende2002, Meunier2017b}. More recently, \citet{Liebing2021}  developed a new, robust approach for measuring convective blueshift. With HARPS spectra of 810 FGKM main sequence stars \citep{Mayor2003}, they obtain the third signature of granulation relative to the Sun. Furthermore, they use an expanded list of lines and provide expected blueshift velocities (observed convective blueshift) for each spectral subtype in their sample. 

Predicting convective blueshift will allow us to select host stars with low granulation RV noise and improve the detection and characterization of low-mass, long-period exoplanets. 
The strength of convective blueshift depends on the fundamental stellar parameters such as effective stellar temperature, surface gravity, and metallicity \citeg{Beeck2013, Meunier2017a, Miklos2020, Diaz2022}. The granulation pattern varies on the timescales of a few minutes and introduces RV variations on the order of a few \ms, which is an order of magnitude larger than the expected amplitude of an Earth twin \citep{Cegla2019}. Furthermore, the magnetic fields on the Sun affect the granulation patterns visible on its surface. In plages, where the magnetic field is more intense and concentrated, the granules are smaller than in the quiet Sun \citep[see][]{Title1992, Narayan2010}. Stellar granulation also significantly impacts the detectability of low-mass planets at longer periods \citep{Meunier2020}. 


In this paper, we obtain the amplitude of convective blueshift for 320 main sequence F, G and K stars using semi-empirical stellar models and estimate the dispersion in radial velocity due to granulation. We begin with details of our target selection in Sect. \ref{sec:targetselection}.  We describe our semi-empirical model for convective blueshift and give a relation to determine it for different stellar parameters in Sect. \ref{sec:model}. We compare our modelled convective blueshift with observed convective blueshift from \citet{Liebing2021} in Sect. \ref{sec:comparison}. We compute the noise due to granulation in the radial-velocity data in Sect. \ref{sec:rvnoise}. We summarize our results and conclude in Sect. \ref{sec:result}.

\section{Data}\label{sec:targetselection}
We obtain the sample of stars and their stellar parameters for our analysis from previous works, which we reference and discuss below.

\subsection{Description of the Sample}

The targets in our study are a subset of the sample of stars in \citet{Liebing2021}, hereafter L21. The L21 sample is derived from 3094 stars from the HARPS database \citep{Trifonov2020}. L21 remove 458 subgiants for future work and compute the convective blueshift from the spectra of only main-sequence stars. L21 exclude fast-rotating targets as their method works best for targets with a projected rotational velocity of $(v \, \sin i)$ < 8 \kms, given the limitation set by the HARPS resolving power. Finally, L21 remove 439 targets that have no matching counterpart in \textsc{Gaia} DR2 \citep{Gaiacoll2018}, which yields L21's final sample of 810 stars spanning from F to M spectral types.

For our sample, we remove the M-dwarfs due to their intrinsically higher level of magnetic activity that is known to change the properties of the convective layer in dwarf stars \citep{Title1987,Hanslmeier1991}. 
To identify and remove M-dwarfs from our sample, we utilized L21 spectral classification (see Table E.2 in \citet{Liebing2021}).
We exclude two G-type stars for which the observed convective blueshift measured by L21 departs significantly from the general trend\footnote{We considered outliers to have convective blueshift relative to the Sun smaller than $-4$.}.
\begin{figure}
   \centering
   \includegraphics[scale=0.20]{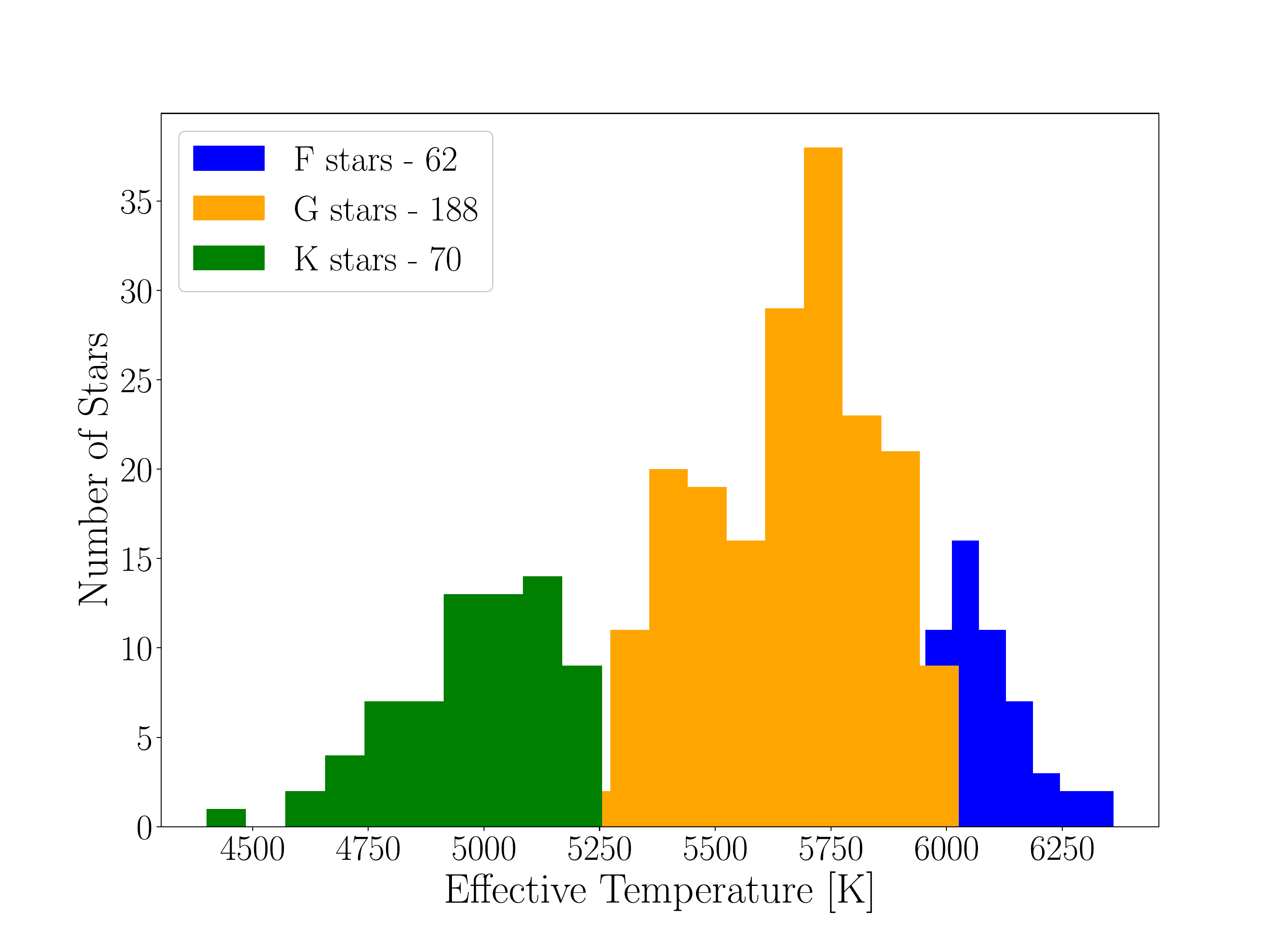}
   \caption{\textit{Distribution of spectral types:} The figure shows the histogram of the number of stars per spectral type within our sample of 320 stars. The blue distribution represents 62 F stars, the orange distribution represents 188 G stars, and the green distribution represents 70 K stars.}  \label{Figspectraltype}       
\end{figure}
   
Furthermore, we use the SWEET-Cat\footnote{\href{http://sweetcat.iastro.pt}{http://sweetcat.iastro.pt}} stellar parameters as they provide the most precise and accurate values for HARPS stars due to their homogeneous analysis \citep{Santos2013, DelgadoMena2017}. \citet{DelgadoMena2017} derive the effective stellar temperature (\teff) and stellar gravity (\logg) for 320 stars out of 738 stars in our sample. This reduces our sample to 62 F stars, 188 G stars, and 70 K stars. Hence, our final sample contains 320 stars in total. Figure \ref{Figspectraltype} shows the distribution of spectral types for our sample. 
Similarly to the L21 sample, our sample is dominated by G-type stars.

\subsection{Stellar Parameters}

We obtain the stellar temperature, corrected surface gravity and metallicity from Table 1 in \citet{DelgadoMena2017}. The stellar parameters of stars with \teff $> 5200 K$ in \citet{DelgadoMena2017} are initially derived by \citet{Sousa2008, Sousa2011a, Sousa2011b}. However, \citet{DelgadoMena2017} re-derive the parameters for stars with \teff $< 5200 K$ using the line list present in \citet{Tsantaki2013}. In particular, the spectroscopic \logg\ is not comparable to the trigonometric \logg\ or the asteroseismic \logg. Various studies apply a correction which is dependent on the effective temperature to derive spectroscopic \logg\ \citep[e.g.,][]{Tsantaki2013, Mortier2014}. \citet{DelgadoMena2017} also derive this correction by fitting a linear trend between the spectroscopic \logg\ and \teff\, corresponding to different spectral types. They apply the correction to the \logg\ values using Equations 1, 2 and 3 from their paper.  

We use stellar mass estimates from Table 2 in \citet{DelgadoMena2019}. To derive the stellar mass, they utilize values for \teff and metallicity from \citet{DelgadoMena2017}, and \textit{V} magnitudes from the HIPPARCOS catalogue \citep{Perryman1997}, while the parallaxes are taken from Gaia DR2  \citep{Gaiacoll2018,Lindegren2018}. \citet{DelgadoMena2017} employ the \textsc{param} v1.3 tool, by using \textsc{parsec} isochrones \citep{Bressan2012} and a Bayesian estimation method, as presented in \citet{Silva2006} to determine the stellar mass.

Finally, to derive the radius of the stars, we use photo-geometric distances from \citet{BailerJones2021} and apparent magnitudes from \textsc{gaia} DR3 \citep{gaia2022}. No correction is needed for interstellar reddening since all stars are in the solar neighbourhood. It is known that there is little or no extinction due to interstellar medium, \textit{E(B$-$V)} $< 0.03$, for stars within 50-100 parsec of our solar system \citep{whittet2018dust}. We also run a simple Monte Carlo analysis for $10^5$ samples to obtain the uncertainty on the radius (the complete procedure for deriving radius and its uncertainty is described in appendix~\ref{app:errR}). 
A complete list of stars with their parameters can be found in Table \ref{tab:stellarparam} (full version available in electronic form). 

\section{Model} \label{sec:model}
We build a semi-empirical model to predict the RV rms due to granulation ($\sigma_{g}$) for all the stars in our sample. We adopt a model where we consider a spherical host star with $N$ granular cells on the visible disk. The hot plasma inside the granules rises with a velocity $v_c$ (or, convection velocity) and falls into the intergranular lanes with the same velocity \citep{Oba2017}.
Since there is a temperature difference between the granule and the intergranular lane, it leads to an intensity contrast between them. We derive a simple relation for $\sigma_{g}$ that depends on the stellar parameters, subject to several simplifying assumptions. The assumptions we make are intrinsic to mixing length theory, which explains the typical behaviour of convective motions by analyzing how they change on a characteristic length scale, the mixing length \citep{Gough1976}. 

To find the dependence of $\sigma_{g}$ on stellar parameters, we need first to find the dependence of the number of granules and convective blueshift on such stellar parameters, which we do below.

\subsection{Number of granules}
The total number of granules, $N$,  on the visible surface of the star (ignoring foreshortening) can be approximated to:
\begin{equation}\label{eqnumberofgranules}
    N \;\;  \propto  \;\;  \left(\frac{R}{r}\right)^2 \ ,
\end{equation}
where $R$ is the stellar radius, and $r$ is the typical size of a granule. As in \citeauthor{BasuChaplin2017} (\citeyear{BasuChaplin2017}, hereafter BC17), we consider that $r$ scales linearly with the mixing length $l_c = \alpha H_P$. We also assume that $\alpha$, the mixing-length parameter, is the same for all stars, given that our stars are similar. The pressure scale height, $H_P$, is proportional to the effective temperature and inversely proportional to the surface gravity of the stars for a simple isothermal atmosphere approximation. With those assumptions, we write $H_P$ and $g$ as below,
\begin{equation}\label{eq2numberofgranules}
    H_P = \frac{\mathcal{R} T_{\mathrm{eff}} }{ \mu g} \,\,;\,\, g \propto \frac{M}{R^2} \   ,
\end{equation}
where $\mathcal{R}$ is the gas constant and  $\mu$ is the mean molecular weight. We assume $\mu$ to be 1 for our analysis. $M$, $g$ and $T_{\mathrm{eff}}$ are the stellar mass, stellar gravity, and effective surface temperature of the star, respectively. This further allows us to rewrite Eq.~\ref{eqnumberofgranules} as:
\begin{equation}
    N \;\;  \propto  \;\;  \left(\frac{M}{R \,T_{\mathrm{eff}}}\right)^2 \ .
\end{equation}
The number of granules on the surface of a main sequence star increases as its temperature decreases. Consequently, K-type stars exhibit a higher number of granules on their surface than F-type stars.

\subsection{Convective Blueshift $\mathcal{V^{\dagger}}$}
In our model, we assume that the mean variation of the Doppler velocity of a granular cell is given by:
\begin{equation}
\mathcal{V} =  f_c \times v_c \ ,
\end{equation}
where $f_c$, the convection factor, is applied to account for the partial cancellation of signals due to hot (bright) granules and cooler (darker) intergranular lanes, and $v_c$ is the convection velocity. We compute the convective blueshift of the star by integrating over the stellar surface that is visible to the observer (Eq. 4.78 in BC17):
\begin{equation}
 \mathcal{V}^{\dagger} = \frac{2}{\pi}\,\mathcal{V} \ .
\end{equation}

In the following subsections, we describe the convection factor and convection velocity for granular cells of different stellar types.

\subsubsection{Convection velocity, $v_c$: }
We use the convection velocity from Eq.~4.73 in BC17, which depends on the stellar parameters. BC17 assumes that the parcel of gas in the convection zone behaves adiabatically and ignores any changes in the mean molecular weight due to ionization. Another important assumption made by the authors is that convection carries all the stellar flux to the surface. For the complete list of underlying assumptions made while deriving the expression for $v_c$, we reference the reader to Section~4.3.1 from Chapter~4, in BC17 and  \citet[][]{Samadi2013, Samadi2013b}. We thus write the convection velocity as:
\begin{equation}
   v_c \;\;  \propto  \;\; T_{\mathrm{eff}}^{32/9}\;g^{-2/9} \ . 
\end{equation}

\subsubsection{Convection factor, $f_c$: } 
Granules are brighter than the intergranular lanes and cover a broader surface area. These different-sized granules have an intensity that changes with their respective size \citep{Hirzberger1997,Berrilli2002}. Additionally, the intensity contrast of granules changes for different stellar types, which we need to consider while calculating the convective blueshift \citep{Trampedach2013,Beeck2013}. To account for the different sizes and intensities of the granules, we assume that the convection factor is proportional to the square of the intensity contrast of the granules rather than simply considering it to be the same for all stars: 
\begin{equation}\label{eq:fc}
    f_c \propto \mathrm{I}_{\mathrm{c}}^{2} \ ,
\end{equation}
where \Ic\ is the intensity contrast of granules. We use the relation for \Ic\ derived from three-dimensional (3D), convective, stellar atmosphere simulations from \citet{Trampedach2013}. They provide the following relation of \Ic\ as a function of stellar parameters for main-sequence stars satisfying \logg\ $\geqslant$ 4:
\begin{equation}\label{eq:Ic}
   \mathrm{I}_{\mathrm{c}} \;\;  =  \;\; 54.98 \times \log T_{\mathrm{eff}} \,-\,4.80 \times \log g \, -\,169.00 \ .
\end{equation} 

From those relations for $f_c$ and $v_c$, we compute the convective blueshift of our stars with respect to the Sun instead of finding their absolute value because the absolute value is not well constrained by observations for stars other than Sun. Hence, the final expression for our model of convective blueshift with respect to solar convective blueshift is:
\begin{equation}\label{eq:solarCB}
  \mathcal{V^{\dagger}} =  \frac{v_{c,\star}}{v_{c,\odot}} \, \left(\frac{I_{\mathrm{c},\star}}{I_{\mathrm{c},\odot}}\right)^2 \ .
\end{equation}

We propagate the uncertainty on this solar relative convective blueshift $\mathcal{V^{\dagger}}$ using Equations in Appendix \ref{eqCBerr}. Since both $v_c$ and \Ic\ are dependent on $T_{\mathrm{eff}}$ and $g$, it makes our model for convective blueshift sensitive to these stellar parameters and their uncertainties. 

\subsection{Granulation noise}
The radial-velocity rms due to granulation (or granulation noise) is the standard deviation of the distribution of the mean velocity of the granular cell ($\mathcal{V}$). For our model, we assume that all the granules behave in a statistically independent manner and their velocity dispersion scales as their velocity. Thus the radial-velocity rms due to granulation from the stellar surface is given by the standard error on the mean velocity \citep{Altman2005}:
\begin{equation}\label{eq_sigmag}
   \sigma_{g} \;\; \simeq  \;\;  \frac{\mathcal{V}}{\sqrt{N}} \ .
\end{equation}
Our expression to estimate the dispersion due to granulation in radial-velocity variations is relative to the Sun:
\begin{equation}
       \sigma_{g,\star} \;\; =  \;\;  \sigma^{\dagger}_{g} \,\times \,\sigma_{g,\odot} \ ,
\end{equation} 

where $\sigma^{\dagger}_{g}$ is the granulation noise for each star relative to the Sun's and $\sigma_{g,\odot}$ is the granulation noise for the Sun.

\section{Comparison to previous works }\label{sec:comparison}
\begin{figure*}
    \includegraphics[scale=0.36]{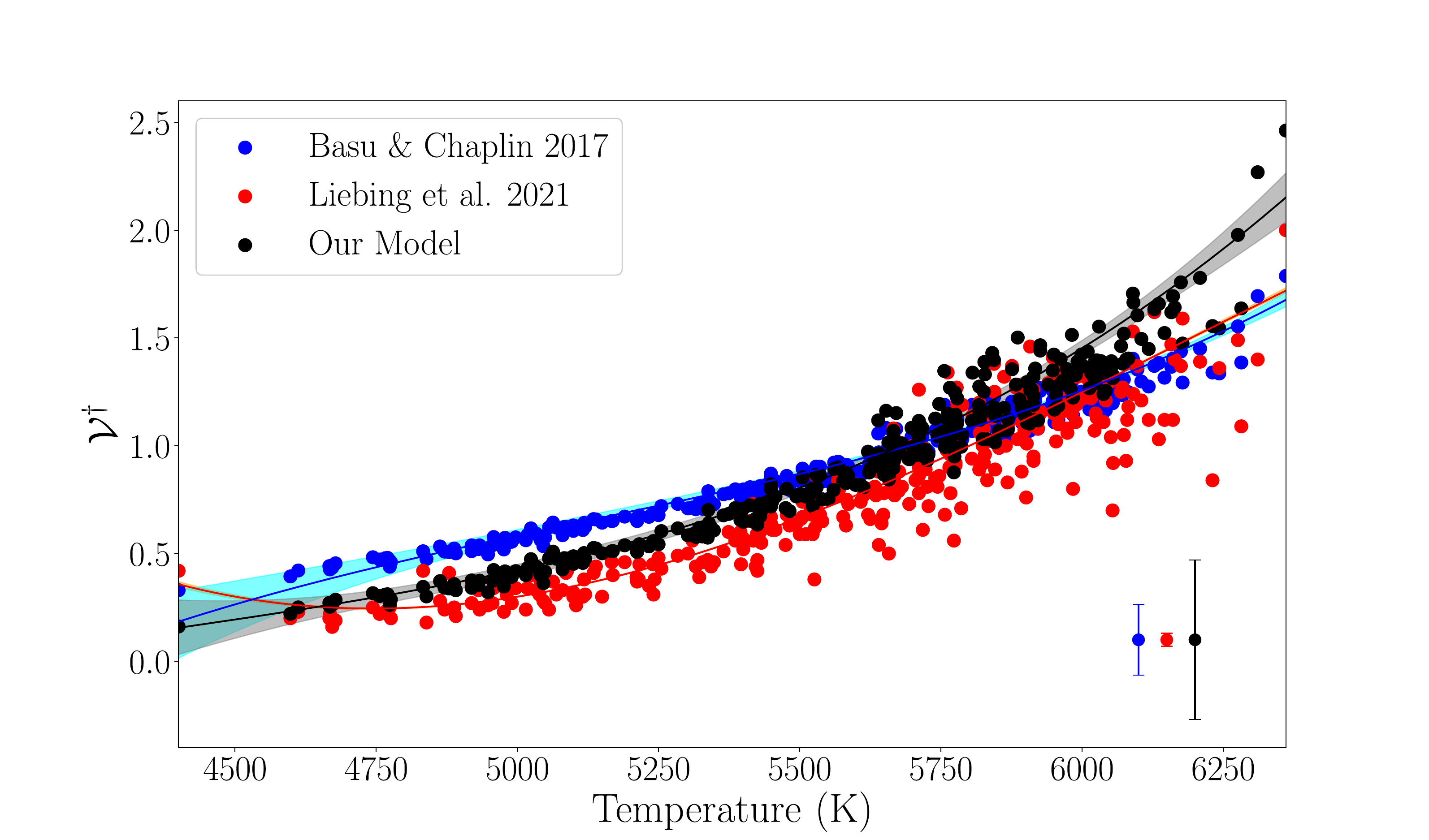}
    \caption{\textit{Convective Blueshift:} In this figure, the  convective blueshift relative to the Sun $ \mathcal{V^{\dagger}}$ is plotted as a function of stellar temperature. The convective blueshift relative to the Sun derived using our model and \citeauthor{BasuChaplin2017} model is shown in solid black and solid blue circles respectively. The observed convective blueshift relative to the Sun from \citet{Liebing2021} is shown in solid red circles. The blue, black, and red solid lines are third-order polynomial fits obtained from the maximum likelihood in the \citeauthor{BasuChaplin2017} model, our newly modelled convective blueshift, and observations, respectively. The shaded cyan, grey, and orange regions are the regions between the 16th and 84th percentiles (i.e., 1-$\sigma$) of the marginal distributions.}\label{figtemplogg}
\end{figure*}

We find the convective blueshift relative to the Sun for all the stars using the model from section \ref{sec:model} and the stellar parameters from section \ref{sec:targetselection}, as shown in figure \ref{figtemplogg}. We compare our model estimates of convective blueshift relative to the Sun with observations from \citet{Liebing2021} and \citeauthor{BasuChaplin2017} model.

\subsection{Comparison with the observed convective blueshift}
We use the solar-relative convective blueshifts estimated by \citet{Liebing2021}, who developed a new approach for measuring convective blueshifts using the third signature of granulation, which involves utilizing the relationship between the depth of spectral lines and their core position. They provided expected blueshift velocities for HARPS spectra of 810 FGKM main sequence stars. The rationale for using their dataset is that it serves as the basis for our sample and thus allows us to compare our modelled convective blueshift with their observed values. Figure \ref{figtemplogg} shows the solar-relative convective blueshift as a function of stellar temperature. The solid red circles are the observed $ \mathcal{V^{\dagger}}$ from \citet{Liebing2021}, while the solid black circles are the derived $ \mathcal{V^{\dagger}}$ using our model from Eq. \ref{eq:solarCB}. 

In L21, the authors derive an empirical relation for the convective blueshift as a function of the stellar temperature by fitting a third-order polynomial to their data, which they median-binned by temperature. We use Bayesian polynomial fits with uniform priors and employ a Markov chain Monte Carlo (MCMC) routine in order to recover the full dataset information, which would otherwise be lost by binning the data. We test not only 3rd-order polynomials but also 2nd-order polynomials through a Bayesian procedure in order to encompass the different behaviour expected from the lack of M-dwarfs and the fact that our dataset is different from the one used in L21. We maximise the likelihood through an MCMC procedure to evaluate if different polynomials could be more appropriate for the observed convective blueshift. We also examine the different polynomials for the complete L21 sample without M-dwarfs (i.e., 738 stars) as it allows us to investigate if the exclusion of almost 50\% of the stars (320 out of 738 stars) has a significant impact on the derived empirical relation for the convective blueshift as a function of stellar temperature. The results of the analysis are summarised in Table \ref{tab:mcmc_analysis}.

\textbf{Information Criteria - AIC and AICc :} The simplest way to compare the two models would be to compare the log-likelihoods. However, this runs the risk of over-fitting (or under-fitting) the data if too many (or too few) free parameters are used. We consider the \emph{corrected} Akaike Information Criterion (AICc, see Appendix \ref{app:aic} and \citealt{Sugiura1978,Burnham&Anderson02}) to distinguish between our different models. The tested model is less appropriate than the reference one with a 95\% confidence if the difference between the AICc of the two models is greater than 6. 

With the L21 sample of 738 stars, we can reject the second-order polynomial fit with  95.9\% confidence and with our sample of 320 stars, we can reject the second-order polynomial fit with 100\% confidence. We confirm that the third-order polynomial best describes the observed convective blueshift with a lower AICc. Therefore, we perform a similar Bayesian analysis for our 320 targets by fitting 3rd-order polynomials in observed convective blueshift, and convective blueshift derived using our model and the \citeauthor{BasuChaplin2017} model.  In Figure~\ref{figtemplogg}, we show the third-order polynomial fits obtained from the maximum likelihood for the \citeauthor{BasuChaplin2017} model (blue line), our model (black line), and L21 observations (red lines). The shaded cyan, grey, and orange regions  are the regions between the 16th and 84th percentiles (i.e., 1-$\sigma$) of the marginal distributions. The results of these fits are given in Table \ref{tab:mcmc_analysis} and highlighted in different colours.

\textbf{Uncertainty on  $\mathcal{V^{\dagger}}$:} It is crucial to emphasize that the uncertainty on all datasets is taken into consideration during the polynomial fitting and comparison processes. The uncertainty in both our model and the model by \citeauthor{BasuChaplin2017} for convective blueshift increases with \teff, as shown in equation \ref{eqCBerr}. Assuming that the uncertainties provided by \citet{Liebing2021} underestimated the errors\footnote{In \citet{Liebing2021}, the uncertainties are computed from the fitting algorithm, which is calculated from the statistics of the fit, rather than being propagated from other parameters.}, given the significant differences in the error bars of three datasets shown in Fig. 3, we test the robustness of our conclusions by assuming errors on the observed convective blueshift that are twice as large. Still, the conclusions remain consistent up to a 1-$\sigma$ level.

The observed and modelled convective blueshift estimates match remarkably well given the simplicity of our model across the full range of effective temperatures probed. The main discrepancy between the model and observations is a relatively constant offset, which may be slightly increasing at higher temperatures. 
To quantify the offset between our model and the observations, we subtract the two polynomial fits (red and black) and take the mean value for this offset over the entire temperature range. We find this mean offset for all the stars to be 0.16 $\pm$ 0.06 (relative to the Sun), which remains within the uncertainties of the modelled convective blueshift (i.e., $\sim$ 0.4). Therefore, considering the given uncertainties, our model demonstrates the ability to accurately forecast the convective blueshift for these stars.

Several previous works have observed this type of offset, particularly while comparing observed convective blueshift with simulations/models (as in \citealt{Meunier2017a}), but the reason behind it is not clear. We hypothesize that using the Sun as a reference when retrieving the observed convective blueshift may lead to an offset. Additionally, there are a few caveats to our model which could explain the discrepancy (an offset) between the observed and our modelled convective blueshift. For instance, the assumptions about ignoring foreshortening effects while computing the number of granules and convective blueshift could create such an offset. Depending on the position of granules on the disk, they do not have the same contribution to the total ``signal'' (RV or intensity). In our work, we assume that the size of granules remains constant from the centre to the limb of the stellar disk. However, observations of granulation by \citet{Johannes2018} showed that the blueshift decreases and can even turn into a redshift toward the solar limb. We expect that including this effect in our model will reduce the value of our modelled blueshift, eventually diminishing the offset between our model and observations. Another caveat to our model is that while computing the convective blueshift, we do not consider any wavelength dependence on the velocity and intensity contrast of the granules. Since spectral lines are formed at different heights in the atmosphere, they have different Doppler shift velocities and line contrasts \citep[such as][]{Reiners2016,Liebing2021}. 

Furthermore, the value of intensity contrast of granules used in our model is derived using an equation that does not consider the metallicity dependence and is valid only for solar metallicity. This results in less scatter around the fit for our model compared to observations. Many previous works \citep[e.g.,][]{AllendePrieto2013, Magic2013,  Corsaro2017, Tayar2019} have found that the convective blueshift, as well as granulation noise in the radial-velocity measurements, has a subtle dependence on the stellar metallicity.

The additional scatter in the observed convective blueshifts compared to the modelled convective blueshifts among stars of similar spectral type may be attributed to differences in activity levels or observation times \citep{Liebing2021}. This is consistent with the findings of \citet{Meunier2017a,Meunier2017b}, who reported an anticorrelation between convective blueshift and activity. Further research into the relationship between stellar variability and convective blueshift could shed light on the underlying mechanisms driving these phenomena.



\subsection{Comparison with the Basu and Chaplin model}
\cite{BasuChaplin2017} present a mathematical approach to model convective blueshift velocity amplitudes using mixing-length theory and establish scaling relations (Eq. 4.73 in BC17). In their model, the authors assume $f_c$ to be 1. In contrast, we assume $f_c \propto \mathrm{I}_{\mathrm{c}}^{2}$ (see Eqs.~\ref{eq:fc} and \ref{eq:Ic}), implying that the intensity contrast of granules changes with spectral type. We compare our modelled convective blueshift values with those derived using the model from \cite{BasuChaplin2017}. In Figure~\ref{figtemplogg}, we present a comparison of the convective blueshift relative to the Sun, as derived using our model from Eq. \ref{eq:solarCB} (represented by solid black circles) and the model by \citeauthor{BasuChaplin2017} (represented by solid blue circles). The figure also displays the third-order polynomial fits obtained from the maximum likelihood for both the \citeauthor{BasuChaplin2017} model (blue line) and our model (black line).

We perform a Kolmogorov–Smirnov \citep[][hereafter KS]{Kolmogorov1933, Smirnov1939} test to quantify the agreement between the distribution of observed convective blueshift and modelled convective blueshift for both our model and the \citeauthor{BasuChaplin2017} model. 
The KS test is used to compare two 1-D distributions in order to determine whether or not they come from the same parent distribution. As this test is sensitive to shifts between the two compared distributions, we adapt it to our purposes (i.e., comparing two offset distributions) by subtracting the medians of the \citeauthor{BasuChaplin2017} model, observations, and our model from their distributions, respectively. We remind the reader that a lower KS statistic indicates a better agreement between the two compared distributions. 

We find the value of the KS statistic to be 0.181 (p-value = $5.2 \times 10^{-5}$) for the \citeauthor{BasuChaplin2017} model and observations. The KS test for our model and observations yields a KS statistic of 0.062 (p-value = 0.56). This implies that our model agrees significantly better with observations than the \citeauthor{BasuChaplin2017} model. This improvement is visible in Figure~\ref{figtemplogg}, where the \citeauthor{BasuChaplin2017} model strongly overpredicts the convective blueshift at lower temperatures (i.e., \teff $\lesssim 5300$\, K).


\section{Granulation Noise}\label{sec:rvnoise}
\begin{figure}
\centering
\includegraphics[width=0.99\hsize]{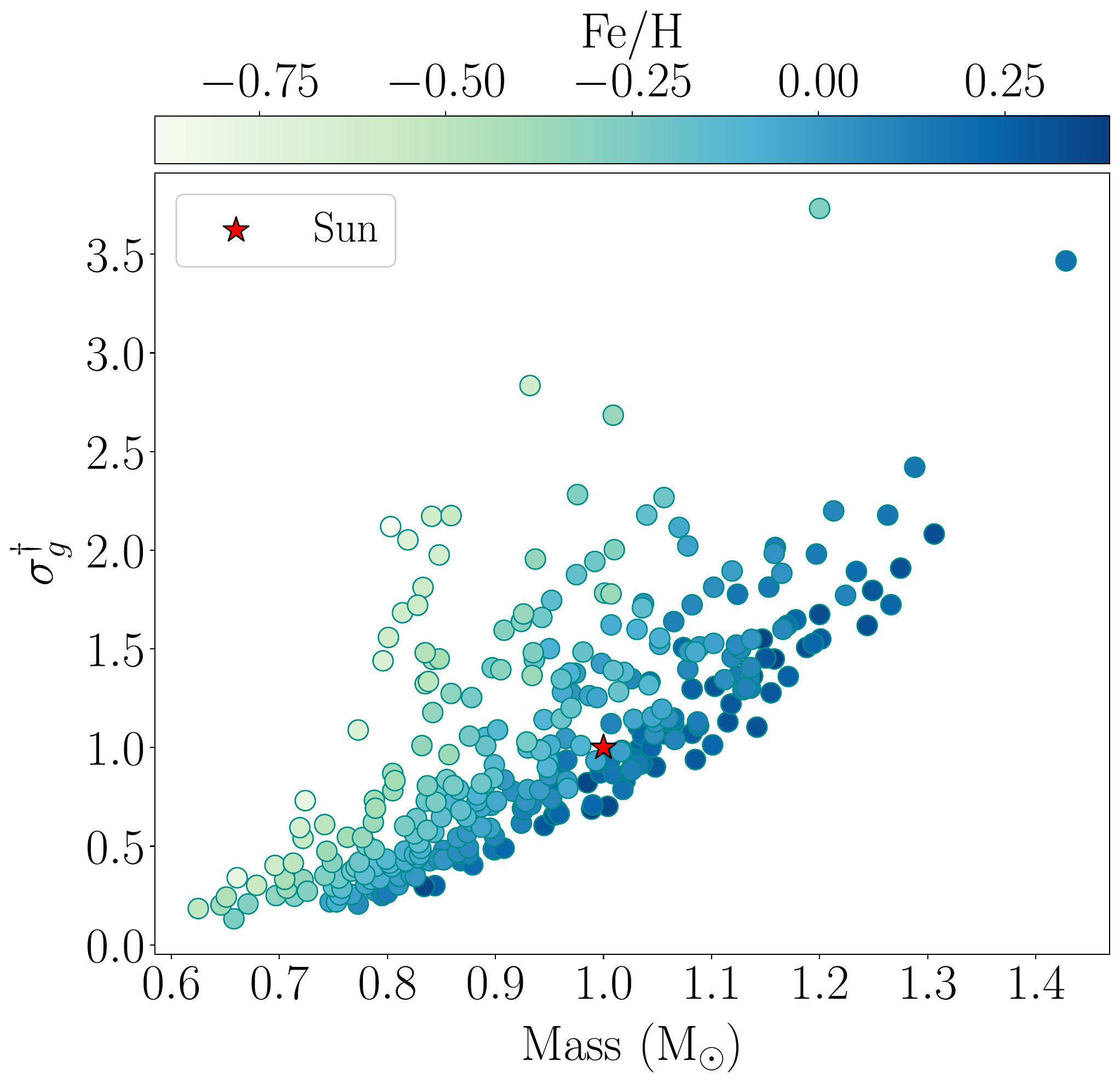}
\caption{\textit{Granulation Noise :} In this figure, the dispersion in the radial velocity due to granulation relative to the Sun, $\sigma^{\dagger}_{g}$, is plotted as a function of stellar mass. For our Sun, the value of $\sigma^{\dagger}_{g}$ is 1. The different colours in the plot represent the metallicity of the host star.}
\label{fig_sigma}%
\end{figure}

\begin{table*} 
\caption{Results for the  Bayesian quadratic fit in different metallicity bins using equation \ref{eq:metalfit}}\label{tab:metallicityfit}
\begin{tabular}{ |c |c| c| c| } 
 \hline
 \rule{0pt}{3ex}   Metallicity bin & $b_{i}$ & $c_{i}$ & $d_{i}$ \\  [2pt]
 \hline\hline
  \rule{0pt}{3ex} [-0.89,-0.57] &-31.927 $_{- 43.919 }^{+ 35.186 } $ & 0.047 $_{- 9.379 }^{+ 7.536 } $ &  0.493 $_{- 0.482 }^{+ 0.388 } $ \\ [2ex]
\hline
 \rule{0pt}{3ex} $[-0.57,-0.25]$ &-18.706 $_{- 12.729 }^{+ 12.526 } $ & 2.262 $_{- 1.836 }^{+ 1.79 } $ &  0.314 $_{- 0.072 }^{+ 0.07 } $ \\ [2ex]
\hline
 \rule{0pt}{3ex} $[-0.25,0.07]$ & -22.569 $_{- 12.978 }^{+ 14.251 } $ & 2.721 $_{- 0.871 }^{+ 0.962 } $ & 0.125 $_{- 0.034 }^{+ 0.032 } $ \\ [2ex]
\hline
 \rule{0pt}{3ex} $[0.07,0.39]$ & -6.339 $_{- 14.778 }^{+ 15.469 } $ & 4.133 $_{- 1.319 }^{+ 1.227 } $ & -0.06 $_{- 0.055 }^{+ 0.057 } $ \\ [2ex]

\hline

 \end{tabular}
 \end{table*}
We compute the RV rms due to granulation by using equation \ref{eq_sigmag} for the stars in our sample. Figure \ref{fig_sigma} shows the granulation RV noise relative to the Sun as a function of stellar mass. The colour scale denotes the metallicity of the star. 
We find that stars with higher metallicities display lower radial-velocity dispersion due to granulation for a given mass range. This is because the strongest blueshifts arise from lines that form in the deeper atmospheric layers, where the convection is most vigorous. The increased continuum opacity at higher metallicity makes these deeper layers less visible. \citet{Meunier2017b} also finds an inverse correlation between convective blueshift and metallicity for stars in the F7-K4 range.

\begin{figure}
\centering
\includegraphics[width=0.99\hsize]{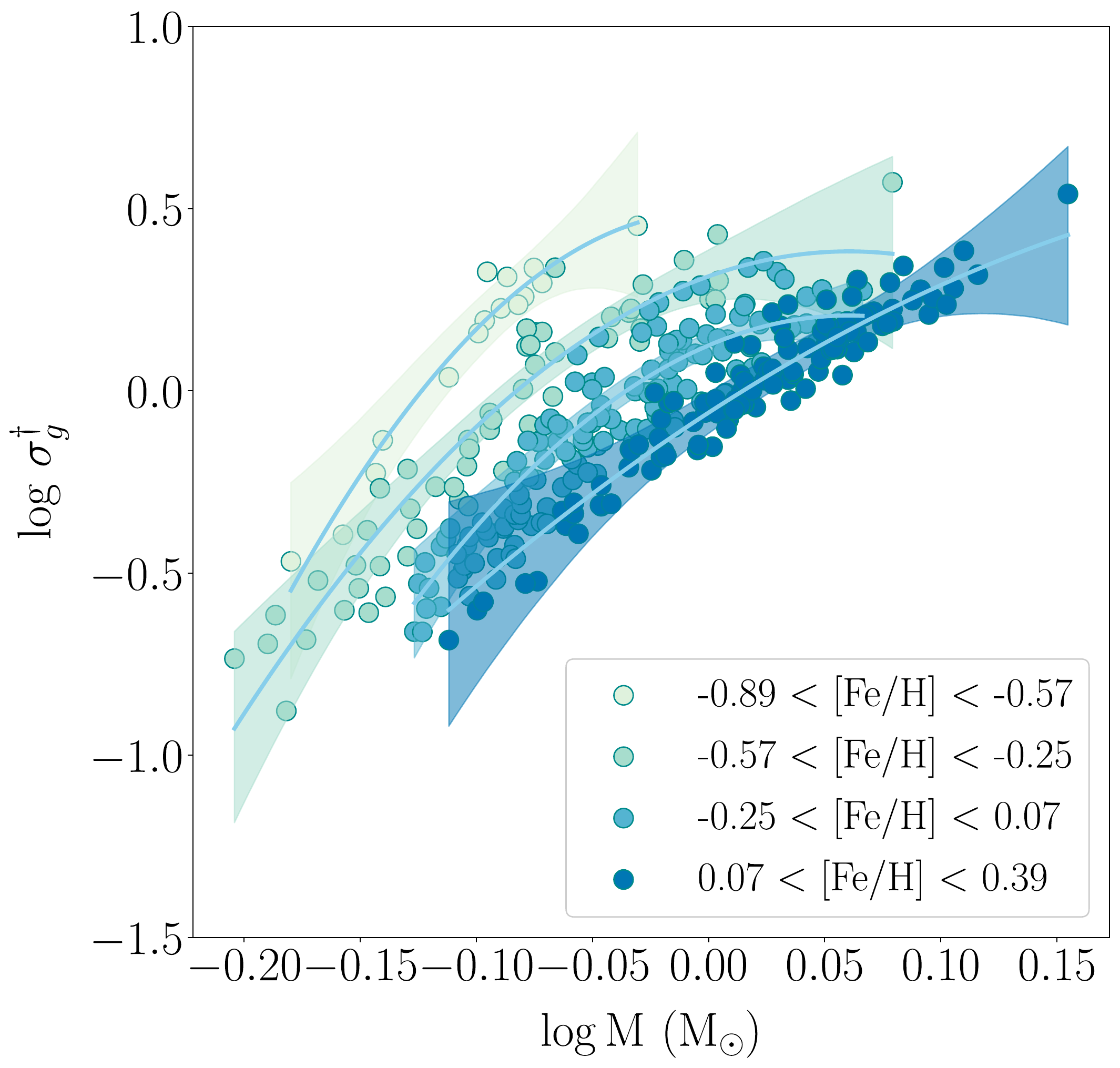}
\caption{\textit{Granulation Noise with metallicity:} This figure is similar to Figure \ref{fig_sigma} except that both of the axes are in the log scale. The different colours in the plot represent the various bins of metallicity for the stars in our sample. The solid sky blue lines are the Bayesian quadratic fits in each metallicity bin.}
\label{fig_sigmafit}%
\end{figure}

We fit empirical relations that the reader can use to find the granulation noise for a large number of stars with stellar temperatures between 4400 K and 6300 K. We use a Bayesian quadratic fit with uniform priors and employ an MCMC routine to fit the $\log \sigma^{\dagger}_{g}$ and $\log M$ for all the stars in each metallicity bin. We use the following quadratic law for the four metallicity bins:
\begin{equation}\label{eq:metalfit}
\log \sigma^{\dagger}_{g} = b_{i} (\log \mathrm{M})^{2} + c_{i} (\log \mathrm{M}) + d_{i} ,
\end{equation}
where $b_i$, $ c_i $, and $d_i$ are the coefficients at maximum likelihood, their uncertainties are from the 16th and 84th percentiles of the marginal distributions. The results of these fits are given in Table \ref{tab:metallicityfit}. We can use the above equation to quantify the radial-velocity rms due to granulation for various stellar types.

To this point, our analysis has been done with respect to the Sun. Here, we estimate the value of convective blueshift for the Sun using the following realistic assumptions: we assume the convective blueshift of our Sun to be 300 \ms\ \citep{Dravins1999} and the number of granules to be 4 million (2 million on the visible disk). We find the dispersion in radial velocity due to granulation to be 0.33 \ms. This value is in agreement with several simulations and observations which found a range of typical rms of the radial-velocity residuals from 0.3 \ms\ \citep[e.g.,][]{Palle1999, Cameron2019, Sulis2020, Moulla2023} to 0.8 \ms\ \citep[e.g.,][]{Meunier2015}. This agreement leads to further confidence in our model and predictions of $\sigma^{\dagger}_{g}$. Assuming the above value for solar granulation noise, we find the dispersion in radial velocity due to granulation for stars in our sample to range from 5 \cms\ (for K-dwarfs) to 80 \cms\ (for F stars). This indicates that K stars are the most suitable targets for monitoring when searching for Earth-mass planets.


\section{Conclusion} \label{sec:result}
The impact of intrinsic stellar noise becomes significant when detecting and characterising smaller exoplanets for both the transit and the Doppler method. In this work, we find an expression for the stellar noise induced by granulation noise based on stellar parameters (i.e., \teff, \logg, $M_{\star}$, $R_{\star}$) for solar-type stars (see Equation \ref{eq:metalfit}). We have analysed a sample of 320 main-sequence FGK stars with \teff between 4400 K and 6300 K and derived the amplitude of the convective blueshift for each star using our semi-empirical stellar model. We also compared our results with the convective blueshift obtained with observations. The results from our analysis are summarised below:

 \begin{itemize}
     \item Stellar convective blueshift increases with the effective temperature of the star. The convection of cooler stars is less visible than hotter stars because it occurs beneath an optically thick layer. This is evident in the predicted convective blueshift of cooler stars, which is lower than that of hotter stars. We also fit a third-order polynomial to both observed and modelled convective blueshift as in \citet{Liebing2021} and provide the results of our fits in Table~\ref{tab:mcmc_analysis}. 
     \item Our modelled convective blueshift agrees well with the observed convective blueshift of \citet{Liebing2021} and it exhibits considerable improvements over the \citeauthor{BasuChaplin2017} model, as discussed in Section 4. We observe an offset  ($\sim$ 0.16*300 = 48 \ms) between the observed and our modelled convective blueshifts. This offset has been observed in previous works \citep{Meunier2017a} but needs better understanding.
     \item We compute the radial-velocity dispersion due to granulation ($\sigma_{g,\star}$) and find that it increases as a function of stellar mass as shown in Figure \ref{fig_sigma}. We find a metallicity dependence on the RV dispersion due to granulation. Low-metallicity stars have higher $\sigma_{g,\star}$ for a fixed stellar mass. This has also been previously found in simulations of \citet{AllendePrieto2013}, where the authors conclude that convective blueshift is enhanced at low metallicity, which implies higher granulation noise.
     \item We provide an empirical relation (equation \ref{eq:metalfit}) to predict the granulation noise ($\sigma^{\dagger}_{g}$) for different star types.  
 \end{itemize}
 
By accurately predicting the granulation noise floor, we can identify host stars with minimal granulation RV noise. This will enhance our ability to detect and characterize low-mass exoplanets with long orbital periods. Our current model indicates that K dwarfs (less massive stars) with higher metallicity are the sweet spot for detecting these low-mass planets. One can use the provided empirical relations to predict the granulation noise that can inform upcoming exoplanet surveys, such as the Terra Hunting Experiment and RV follow-up of PLATO candidates, for selecting the most amenable targets. In a later paper, we will improve our model to have foreshortening effects and metallicity dependence. Furthermore, we will include spots and plages in our model to obtain the suppression of convective blueshift.

\section*{Acknowledgements}
S.D. is funded by the UK Science and Technology Facilities Council (grant number ST/V004735/1). R.D.H. is funded by the UK Science and Technology Facilities Council (STFC)'s Ernest Rutherford Fellowship (grant number ST/V004735/1). W.J.C. acknowledges support from STFC grant ST/V000519/1.
We would like to thank Prof Tim Naylor, Prof Matthew Browning, Prof Isabelle Baraffe, Dr Dimitar G. Vlaykov, Dr Heather Cegla and Dr Eduardo Vitral for many valuable discussions.
We also acknowledge the interesting suggestions from Prof Suzanne Aigrain, Haochuan Yu, Prof Andrew Collier Cameron, and Dr Michaël Cretignier in developing this work.

\section*{Data Availability}

The data used to create the plots presented in this paper and any additional findings discovered in this study can be obtained by contacting the corresponding author and requesting access to the information. The full version of table \ref{tab:stellarparam} can be obtained in electronic form.



\bibliographystyle{mnras}
\bibliography{example} 



\newpage
\appendix

\section{Radius and uncertainity}\label{app:errR}
To compute the radius of a star, we first compute the absolute magnitude from the photo-geometric distances from \citet{BailerJones2021} and apparent magnitudes from \textsc{gaia} DR3 \citep{gaia2022} using the equation below: 
\begin{equation}
        \mathrm{M_G} =\mathrm{m}_{\mathrm{G}} - 5(\log d) + 5 - \mathrm{Ag} \ ,
\end{equation}
where $\mathrm{M_G}$ is the absolute magnitude in the G-band, $\mathrm{m}_{\mathrm{G}}$ is the apparent magnitude and d is the distance of the star from Earth, in parsecs. We set Ag to be 0 for all of our stars \citep{Andrae2018}. The next step is to calculate the luminosity ($\mathrm{L}$) of the star in units of solar luminosity. We use the bolometric correction (BC), which is dependent on the effective temperature of the star, using the eq. 7 from \citet{Andrae2018}. We compute the luminosity with
\begin{equation}
-2.5 \log \mathrm{L} =  \mathrm{M_G} + BC - BC_{\odot} \ ,
\end{equation}
where $BC_{\odot}$ = 4.74 mag is the solar bolometric magnitude, as defined in IAU Resolution 2015 B21\footnote{\href{https://www.iau.org/static/resolutions/IAU2015_English.pdf}{IAU Resolution 2015 B21}.}. Finally, with the luminosity of the star, we can find the stellar radius using the equations below:
\begin{equation} \label{eq: rad}
R = \sqrt{\frac{\mathrm{L}\,\mathrm{L_{\odot}}}{4\,\pi\, \sigma\, T_\mathrm{eff}^4}} \ ,
\end{equation}
where $\sigma$ is the  Stefan–Boltzmann constant. 

We employ a Monte Carlo method of error propagation for computing the uncertainty on the stellar radius, by assuming Gaussian errors. In short, we sample $10^5$ Gaussian variables centred in the measured values of distance, apparent magnitude, and temperature, with respective widths given by their nominal uncertainties. Next, we use eq.~\ref{eq: rad} to estimate the radius from these random Gaussian values by taking their mean of the final distribution. The distribution of the calculated values of the radius shows the effect of the imprecision of the input data. Hence, we extract the standard deviation of the final distribution, which we use as the 1-sigma uncertainty on the radius.

\section{Uncertainty on  $\mathcal{V^{\dagger}}$ } \label{appuncertainity}
We propagate the errors from $T_{\mathrm{eff}}$ and $g$ to $v_c$ and \Ic\ as follows:
\begin{equation}\label{eqverr}
\centering
 \sigma_{v_c} = v_{c,\star}\,\, \sqrt{ \left(\frac{32 \, \sigma_{T_{\mathrm{eff}}}}{9 \,T_{\mathrm{eff}}}\right)^2  +  \left(\frac{2 \sigma_{g} }{9 \,g}\right)^2  } ,
 \end{equation}
 
 \begin{equation}\label{eqierr}
  \sigma_{\mathrm{I}_{\mathrm{c}}} = \sqrt{ \left(\frac{54.98 \,\sigma_{T_{\mathrm{eff}}}}{T_{\mathrm{eff}} \,\ln{10} }\right)^2  +  \left(\frac{4.80  \, \sigma_{g} }{g\,\ln{10} }\right)^2  } ,
\end{equation}
The uncertainty on $\mathcal{V^{\dagger}}$  is given by:

 \begin{equation}\label{eqCBerr}
\sigma_{\mathcal{V^{\dagger}}} =  \mathcal{V^{\dagger}} \, \, \sqrt{ \left(\frac{\sigma_{v_c}}{v_{c,\star}}\right)^2 + 2 \left(\frac{  \sigma_{ \mathrm{I}_{\mathrm{c}}}}{ \mathrm{{I}_{\mathrm{c,\star}}}}\right)^2  } \ .
\end{equation}

\section{AIC}\label{app:aic}
To distinguish between the two models, we considered the corrected Akaike Information Criterion \citep[AICc][]{Sugiura1978} and the original Akaike Information Criterion \citep[AIC][]{Akaike83} :
\begin{equation}
    \mathrm{AIC} = -2 \ln \mathcal{L} + 2 * N_{\mathrm{free}} \ ,
\end{equation}
where $\mathcal{L}$ is the maximum likelihood found after running the  Markov chain Monte Carlo (MCMC) analysis to explore the parameter space. 
\begin{equation}
    \mathrm{AICc} =  \mathrm{AIC} + 2 \left(\frac{N_{\mathrm{free}} (1 + N_{\mathrm{free}}  ) }{N_{\mathrm{data}} - N_{\mathrm{free}} +1 }  \right)
\end{equation}
where $N_{\mathrm{data}}$ the number of data points. 

\section{Some Additional Material}

\begin{table*}
\centering
\caption{Results for the fit of n-order polynomials ($\sum_{i=0}^{n=2,3} a_{i} x^{i}$). The values of the coefficients of n-order polynomials are at maximum likelihood and the uncertainties on them are from the 16th and 84th percentiles of the marginal distributions. The highlighted rows are the results of 3rd-order polynomial fits in observed convective blueshift and convective blueshift derived using our model and \citeauthor{BasuChaplin2017} model as shown in figure \ref{figtemplogg}. }
\label{tab:mcmc_analysis}
\begin{tabular}{|l|l| c c c c |l|}
\hline
 \rule{0pt}{3ex}Convective Blueshift &  No. of stars  & $a_3$ & $a_2$   & $a_1$    &    $a_0$  & AICc \\\hline\hline
  \rule{0pt}{3ex}Priors   & Uniform  & [-100,100] &[-100,100]   & [-100,100]    &   [-100,100]   &  \\\hline\hline

 \rule{0pt}{3ex}Observation (L21) & 738 & 7.802 $_{- 1.047 }^{ + 1.034 } $ & -0.003 $_{- 0.001 }^{ +0.001 } $ & 1.6 $_{- 1.2 }^{ + 1.1}\, \mathrm{e}^{-7} $  & 21.4 $_{- 7.2}^{ + 7.4 }\, \mathrm{e}^{-12} $ & 18039.818 \\

  \rule{0pt}{3ex}Observation (L21) & 738 & - & 10.829 $_{- 0.118 }^{+ 0.142 } $ & -45.9 $_{- 0.5 }^{ + 0.5 } \, \mathrm{e}^{-4} $ & 49.9 $_{- 41.9 }^{ + 50.4 }\,\mathrm{e}^{-8} $ & 18046.244 \\ [2pt]\rowcolor{gray}

\rule{0pt}{3ex}Observation (L21) & 320 & 41.468 $_{- 2.212 }^{ + 2.545 } $ & -0.022 $_{- 0.001 }^{ + 0.001 } $ & 37.2$_{- 2.3 }^{ + 2.6 } \, \mathrm{e}^{-7}$ & -19.9 $_{- 1.6 }^{ + 1.4}\, \mathrm{e}^{-11} $ & 3856.665 \\ [1.8pt]
\rule{0pt}{3ex} Observation (L21) & 320 & - & 10.483 $_{- 0.229 }^{+ 0.242 } $ & -45.1 $_{- .9 }^{  + 0.8 }\,\mathrm{e}^{-4} $ & 49.9 $_{- 78.3 }^{ + 83.8 } \,\mathrm{e}^{-8}$ & 4020.774 \\ [2pt]
\hline\rowcolor{orange}
\rule{0pt}{3ex} Our Model & 320 & -12.444 $_{- 24.574 }^{ +26.781 } $ & 0.008 $_{- 0.015 }^{ +0.014 } $ & -1.8 $_{- 2.5 }^{ +2.8 } \,\mathrm{e}^{-6}$ & 1.4 $_{- 1.7 }^{ +1.6}\,\mathrm{e}^{-10} $ & -625.976 \\[2pt] \rowcolor{cyan}
\rule{0pt}{3ex} BC17 Model & 320 & -25.928 $_{- 16.777 }^{ + 21.313 } $ & 0.014 $_{- 0.011 }^{ +0.009 } $ & -2.7 $_{- 1.6 }^{ + 2.1 }\,\mathrm{e}^{-6} $ & 1.8 $_{- 1.2 }^{ +1.0 }\,\mathrm{e}^{-10} $ & -1089.816 \\ [2pt]

\hline
\end{tabular}
\end{table*}
\begin{table*}
\caption{Stellar parameters for our 320 stars, $\mathcal{V^{\dagger}}$, and  $\sigma^{\dagger}_{g}$ . Full version is available in electronic form.} \label{tab:stellarparam} 
\begin{tabular}{|r|r|r|r|r|r|r|r|r|r|r|r|}
\hline
Star & $T_{\mathrm{eff}}$ &  $\sigma_{T}$ & $\log g$ &  $\sigma_{\log g}$ &  [Fe/H] & Mass &  $\sigma_{M} $ &   Radius &  $\sigma_{R}$  & $\mathcal{V^{\dagger}}$ &  $\sigma^{\dagger}_{g}$\\  
\hline 
$-$ &  K & K & dex & dex & &$M_{\odot}$ &   $M_{\odot}$  &   $R_{\odot}$ &   $R_{\odot}$ & - & - \\ \hline\hline
HD85512  &  4400.0  &   45.0  &  4.72  &    0.1  &  -0.26  &  0.658  &  0.024  &   0.706  &    0.014  &      0.162  &   0.132  \\
HIP26542  &  4598.0  &   72.0  &  4.68  &   0.21  &  -0.53  &  0.625  &  0.015  &   0.656  &    0.021  &       0.22  &   0.184 \\
HD21209A  &  4612.0  &   73.0  &  4.57  &   0.19  &  -0.39  &  0.646  &  0.013  &   0.653  &    0.021  &       0.25  &   0.202  \\
HD13789  &  4667.0  &  104.0  &  4.56  &   0.26  &  -0.01  &  0.747  &  0.021  &   0.744  &    0.033  &      0.271  &   0.218  \\ 
HD203413  &  4668.0  &   99.0  &  4.63  &   0.26  &   0.08  &  0.773  &  0.022  &   0.781  &    0.035  &      0.253  &   0.207  \\
HIP22059  &  4672.0  &   82.0  &   4.6  &   0.25  &  -0.28  &  0.671  &  0.017  &   0.656  &    0.022  &      0.262  &   0.208  \\ 
HD210975  &  4678.0  &   64.0  &  4.52  &    0.2  &  -0.45  &  0.651  &  0.015  &   0.681  &    0.018  &      0.286  &   0.242  \\ 
HD208573  &  4744.0  &   79.0  &   4.5  &    0.2  &   0.05  &  0.789  &  0.021  &   0.829  &    0.027  &      0.316  &   0.273  \\
HD63454  &  4756.0  &   77.0  &  4.56  &   0.22  &   0.13  &  0.795  &  0.021  &   0.797  &    0.026  &      0.303  &    0.25  \\ 
 HD12617  &  4766.0  &   86.0  &  4.55  &   0.24  &   0.14  &    0.8  &   0.02  &   0.825  &    0.029  &       0.31  &   0.264  \\ 
HD87521  &  4770.0  &   71.0  &  4.55  &   0.16  &  -0.01  &  0.767  &  0.021  &   0.762  &    0.022  &      0.311  &   0.255  \\
HD103949  &  4774.0  &   80.0  &  4.73  &   0.21  &  -0.07  &  0.753  &   0.02  &   0.757  &    0.025  &      0.262  &   0.218  \\ 
HD114386  &  4774.0  &   59.0  &  4.61  &   0.14  &  -0.09  &  0.756  &  0.019  &   0.783  &     0.02  &      0.295  &   0.253  \\ 
 HD86140  &  4776.0  &   70.0  &  4.64  &   0.18  &  -0.28  &  0.714  &  0.017  &    0.74  &    0.022  &      0.287  &   0.246  \\ 
 HD27894  &  4833.0  &  209.0  &  4.52  &   0.48  &   0.26  &  0.844  &  0.026  &   0.876  &    0.082  &      0.346  &     0.3  \\ 
...  &  ... & ...  &  ... &   ...  & ... &  ...  &  ...  &   ...  &    ... &     ...  &    ...  \\ 
 \hline
 
\end{tabular}

\end{table*}


\bsp	
\label{lastpage}
\end{document}